\documentclass{article}
\usepackage{psfig}
\usepackage{latexsym}
\begin{document}
\newcommand {\ee}{\end{equation}}
\newcommand {\bea}{\begin{eqnarray}}
\newcommand {\eea}{\end{eqnarray}}
\newcommand {\nn}{\nonumber \\}
\newcommand {\Tr}{{\rm Tr\,}}
\newcommand {\tr}{{\rm tr\,}}
\newcommand {\e}{{\rm e}}
\newcommand {\etal}{{\it et al.}}
\newcommand {\m}{\mu}
\newcommand {\n}{\nu}
\newcommand {\pl}{\partial}
\newcommand {\p} {\phi}
\newcommand {\vp}{\varphi}
\newcommand {\vpc}{\varphi_c}
\newcommand {\al}{\alpha}
\newcommand {\be}{\beta}
\newcommand {\ga}{\gamma}
\newcommand {\Ga}{\Gamma}
\newcommand {\x}{\xi}
\newcommand {\ka}{\kappa}
\newcommand {\la}{\lambda}
\newcommand {\La}{\Lambda}
\newcommand {\si}{\sigma}
\newcommand {\Si}{\Sigma}
\newcommand {\Th}{\Theta}
\newcommand {\om}{\omega}
\newcommand {\Om}{\Omega}
\newcommand {\ep}{\epsilon}
\newcommand {\vep}{\varepsilon}
\newcommand {\na}{\nabla}
\newcommand {\del}  {\delta}
\newcommand {\Del}  {\Delta}
\newcommand {\mn}{{\mu\nu}}
\newcommand {\ls}   {{\lambda\sigma}}
\newcommand {\ab}   {{\alpha\beta}}
\newcommand {\gd}   {{\gamma\delta}}
\newcommand {\half}{ {\frac{1}{2}} }
\newcommand {\third}{ {\frac{1}{3}} }
\newcommand {\fourth} {\frac{1}{4} }
\newcommand {\sixth} {\frac{1}{6} }
\newcommand {\sqg} {\sqrt{g}}
\newcommand {\fg}  {\sqrt[4]{g}}
\newcommand {\invfg}  {\frac{1}{\sqrt[4]{g}}}
\newcommand {\sqZ} {\sqrt{Z}}
\newcommand {\sqk} {\sqrt{\kappa}}
\newcommand {\sqt} {\sqrt{t}}
\newcommand {\sql} {\sqrt{l}}
\newcommand {\reg} {\frac{1}{\epsilon}}
\newcommand {\fpisq} {(4\pi)^2}
\newcommand {\Lcal}{{\cal L}}
\newcommand {\Ocal}{{\cal O}}
\newcommand {\Dcal}{{\cal D}}
\newcommand {\Ncal}{{\cal N}}
\newcommand {\Mcal}{{\cal M}}
\newcommand {\scal}{{\cal s}}
\newcommand {\Dvec}{{\hat D}}   
\newcommand {\dvec}{{\vec d}}
\newcommand {\Evec}{{\vec E}}
\newcommand {\Hvec}{{\vec H}}
\newcommand {\Vvec}{{\vec V}}
\newcommand {\rpl}{{\vec \partial}}
\def\overleftarrow#1{\vbox{\ialign{##\crcr
 $\leftarrow$\crcr\noalign{\kern-1pt\nointerlineskip}
 $\hfil\displaystyle{#1}\hfil$\crcr}}}
\def\lpl{{\overleftarrow\partial}}
\newcommand {\Btil}{{\tilde B}}
\newcommand {\ctil}{{\tilde c}}
\newcommand {\dtil}{{\tilde d}}
\newcommand {\Ftil}{{\tilde F}}
\newcommand {\Ktil}  {{\tilde K}}
\newcommand {\Ltil}  {{\tilde L}}
\newcommand {\mtil}{{\tilde m}}
\newcommand {\ttil} {{\tilde t}}
\newcommand {\Qtil}  {{\tilde Q}}
\newcommand {\Rtil}  {{\tilde R}}
\newcommand {\Stil}{{\tilde S}}
\newcommand {\Ztil}{{\tilde Z}}
\newcommand {\altil}{{\tilde \alpha}}
\newcommand {\betil}{{\tilde \beta}}
\newcommand {\etatil} {{\tilde \eta}}
\newcommand {\latil}{{\tilde \lambda}}
\newcommand {\Latil}{{\tilde \Lambda}}
\newcommand {\ptil}{{\tilde \phi}}
\newcommand {\Ptil}{{\tilde \Phi}}
\newcommand {\natil} {{\tilde \nabla}}
\newcommand {\xitil} {{\tilde \xi}}
\newcommand {\Ahat}{{\hat A}}
\newcommand {\ahat}{{\hat a}}
\newcommand {\Rhat}{{\hat R}}
\newcommand {\Shat}{{\hat S}}
\newcommand {\ehat}{{\hat e}}
\newcommand {\mhat}{{\hat m}}
\newcommand {\shat}{{\hat s}}
\newcommand {\Dhat}{{\hat D}}   
\newcommand {\Vhat}{{\hat V}}   
\newcommand {\xhat}{{\hat x}}
\newcommand {\Zhat}{{\hat Z}}
\newcommand {\Gahat}{{\hat \Gamma}}
\newcommand {\Phihat} {{\hat \Phi}}
\newcommand {\phihat} {{\hat \phi}}
\newcommand {\vphat} {{\hat \varphi}}
\newcommand {\nah} {{\hat \nabla}}
\newcommand {\etahat} {{\hat \eta}}
\newcommand {\omhat} {{\hat \omega}}
\newcommand {\psihat} {{\hat \psi}}
\newcommand {\thhat} {{\hat \theta}}
\newcommand {\gh}  {{\hat g}}
\newcommand {\abar}{{\bar a}}
\newcommand {\Abar}{{\bar A}}
\newcommand {\cbar}{{\bar c}}
\newcommand {\bbar}{{\bar b}}
\newcommand {\gbar}{\bar{g}}
\newcommand {\Bbar}{{\bar B}}
\newcommand {\fbar}{{\bar f}}
\newcommand {\Fbar}{{\bar F}}
\newcommand {\kbar}  {{\bar k}}
\newcommand {\Kbar}  {{\bar K}}
\newcommand {\Lbar}  {{\bar L}}
\newcommand {\Qbar}  {{\bar Q}}
\newcommand {\albar}{{\bar \alpha}}
\newcommand {\bebar}{{\bar \beta}}
\newcommand {\labar}{{\bar \lambda}}
\newcommand {\psibar}{{\bar \psi}}
\newcommand {\vpbar}{{\bar \varphi}}
\newcommand {\Psibar}{{\bar \Psi}}
\newcommand {\chibar}{{\bar \chi}}
\newcommand {\sibar}{{\bar \sigma}}
\newcommand {\xibar}{{\bar \xi}}
\newcommand {\thbar}{{\bar \theta}}
\newcommand {\bbartil}{{\tilde {\bar b}}}
\newcommand {\aldot}{{\dot{\alpha}}}
\newcommand {\bedot}{{\dot{\beta}}}
\newcommand {\alp}{{\alpha'}}
\newcommand {\bep}{{\beta'}}
\newcommand {\gap}{{\gamma'}}
\newcommand {\bfZ} {{\bf Z}}
\newcommand {\BFd} {{\bf d}}
\newcommand  {\vz}{{v_0}}
\newcommand  {\ez}{{e_0}}
\newcommand  {\mz}{{m_0}}
\newcommand  {\xf}{{x^5}}
\newcommand  {\yf}{{y^5}}
\newcommand  {\Zt}{{Z$_2$}}
\newcommand {\intfx} {{\int d^4x}}
\newcommand {\intdX} {{\int d^5X}}
\newcommand {\inttx} {{\int d^2x}}
\newcommand {\change} {\leftrightarrow}
\newcommand {\ra} {\rightarrow}
\newcommand {\larrow} {\leftarrow}
\newcommand {\ul}   {\underline}
\newcommand {\pr}   {{\quad .}}
\newcommand {\com}  {{\quad ,}}
\newcommand {\q}    {\quad}
\newcommand {\qq}   {\quad\quad}
\newcommand {\qqq}   {\quad\quad\quad}
\newcommand {\qqqq}   {\quad\quad\quad\quad}
\newcommand {\qqqqq}   {\quad\quad\quad\quad\quad}
\newcommand {\qqqqqq}   {\quad\quad\quad\quad\quad\quad}
\newcommand {\qqqqqqq}   {\quad\quad\quad\quad\quad\quad\quad}
\newcommand {\lb}    {\linebreak}
\newcommand {\nl}    {\newline}

\newcommand {\vs}[1]  { \vspace*{#1 cm} }

\newcommand {\MPL}  {Mod.Phys.Lett.}
\newcommand {\NP}   {Nucl.Phys.}
\newcommand {\PL}   {Phys.Lett.}
\newcommand {\PR}   {Phys.Rev.}
\newcommand {\PRL}   {Phys.Rev.Lett.}
\newcommand {\IJMP}  {Int.Jour.Mod.Phys.}
\newcommand {\CMP}  {Commun.Math.Phys.}
\newcommand {\JMP}  {Jour.Math.Phys.}
\newcommand {\AP}   {Ann.of Phys.}
\newcommand {\PTP}  {Prog.Theor.Phys.}
\newcommand {\NC}   {Nuovo Cim.}
\newcommand {\CQG}  {Class.Quantum.Grav.}


\font\smallr=cmr5
\newcommand {\npl}  {{\frac{n\pi}{l}}}
\newcommand {\mpl}  {{\frac{m\pi}{l}}}
\newcommand {\kpl}  {{\frac{k\pi}{l}}}

\def\ocirc#1{#1^{^{{\hbox{\smallr\llap{o}}}}}}
\def\ogamma{\ocirc{\gamma}{}}
\def\oM{{\buildrel {\hbox{\smallr{o}}} \over M}}
\def\osigma{\ocirc{\sigma}{}}

\def\overleftrightarrow#1{\vbox{\ialign{##\crcr
 $\leftrightarrow$\crcr\noalign{\kern-1pt\nointerlineskip}
 $\hfil\displaystyle{#1}\hfil$\crcr}}}
\def\overnab{{\overleftrightarrow\nabslash}}

\def\va{{a}}
\def\vb{{b}}
\def\vc{{c}}
\def\tilpsi{{\tilde\psi}}
\def\tbpsi{{\tilde{\bar\psi}}}

\def\delL{{\delta_{LL}}}
\def\delG{{\delta_{G}}}
\def\delc{{\delta_{cov}}}

\newcommand {\sqxx}  {\sqrt {x^2+1}}   
\newcommand {\gago}  {\gamma^5}
\newcommand {\Pp}  {P_+}
\newcommand {\Pm}  {P_-}
\newcommand {\GfMp}  {G^{5M}_+}
\newcommand {\GfMpm}  {G^{5M'}_-}
\newcommand {\GfMm}  {G^{5M}_-}
\newcommand {\Omp}  {\Omega_+}    
\newcommand {\Omm}  {\Omega_-}
\def\Aslash{{}\hbox{\hskip2pt\vtop
 {\baselineskip23pt\hbox{}\vskip-24pt\hbox{/}}
 \hskip-11.5pt $A$}}
\def\Rslash{{}\hbox{\hskip2pt\vtop
 {\baselineskip23pt\hbox{}\vskip-24pt\hbox{/}}
 \hskip-11.5pt $R$}}

\def\kslash{
{}\hbox       {\hskip2pt\vtop
                   {\baselineskip23pt\hbox{}\vskip-24pt\hbox{/}}
               \hskip-8.5pt $k$}
           }    
\def\qslash{
{}\hbox       {\hskip2pt\vtop
                   {\baselineskip23pt\hbox{}\vskip-24pt\hbox{/}}
               \hskip-8.5pt $q$}
           }    
\def\dslash{
{}\hbox       {\hskip2pt\vtop
                   {\baselineskip23pt\hbox{}\vskip-24pt\hbox{/}}
               \hskip-8.5pt $\partial$}
           }    
\def\dbslash{{}\hbox{\hskip2pt\vtop
 {\baselineskip23pt\hbox{}\vskip-24pt\hbox{$\backslash$}}
 \hskip-11.5pt $\partial$}}

\def\Kbslash{{}\hbox{\hskip2pt\vtop
 {\baselineskip23pt\hbox{}\vskip-24pt\hbox{$\backslash$}}
 \hskip-11.5pt $K$}}
\def\Ktilbslash{{}\hbox{\hskip2pt\vtop
 {\baselineskip23pt\hbox{}\vskip-24pt\hbox{$\backslash$}}
 \hskip-11.5pt ${\tilde K}$}}
\def\Ltilbslash{{}\hbox{\hskip2pt\vtop
 {\baselineskip23pt\hbox{}\vskip-24pt\hbox{$\backslash$}}
 \hskip-11.5pt ${\tilde L}$}}
\def\Qtilbslash{{}\hbox{\hskip2pt\vtop
 {\baselineskip23pt\hbox{}\vskip-24pt\hbox{$\backslash$}}
 \hskip-11.5pt ${\tilde Q}$}}
\def\Rtilbslash{{}\hbox{\hskip2pt\vtop
 {\baselineskip23pt\hbox{}\vskip-24pt\hbox{$\backslash$}}
 \hskip-11.5pt ${\tilde R}$}}
\def\Kbarbslash{{}\hbox{\hskip2pt\vtop
 {\baselineskip23pt\hbox{}\vskip-24pt\hbox{$\backslash$}}
 \hskip-11.5pt ${\bar K}$}}
\def\Lbarbslash{{}\hbox{\hskip2pt\vtop
 {\baselineskip23pt\hbox{}\vskip-24pt\hbox{$\backslash$}}
 \hskip-11.5pt ${\bar L}$}}
\def\Rbarbslash{{}\hbox{\hskip2pt\vtop
 {\baselineskip23pt\hbox{}\vskip-24pt\hbox{$\backslash$}}
 \hskip-11.5pt ${\bar R}$}}
\def\Qbarbslash{{}\hbox{\hskip2pt\vtop
 {\baselineskip23pt\hbox{}\vskip-24pt\hbox{$\backslash$}}
 \hskip-11.5pt ${\bar Q}$}}
\def\Acalbslash{{}\hbox{\hskip2pt\vtop
 {\baselineskip23pt\hbox{}\vskip-24pt\hbox{$\backslash$}}
 \hskip-11.5pt ${\cal A}$}}

\begin{flushright}
February 2004\\
DAMTP-2003-9\\
US-03-02\\
hep-th/0302029
\end{flushright}

\vspace{0.5cm}

\begin{center}

{\Large\bf 
Brane-Anti-Brane Solution and\\ 
 SUSY Effective Potential\\
in \\
Five Dimensional Mirabelli-Peskin Model
%
}

\vspace{1.5cm}
{\large Shoichi ICHINOSE
         \footnote{
E-mail address:\ ichinose@u-shizuoka-ken.ac.jp
                  }
}\ and\ 
{\large Akihiro MURAYAMA$^\ddag$
         \footnote{
E-mail address:\ edamura@ipc.shizuoka.ac.jp
                  }
}
\vspace{1cm}

{\large 
Laboratory of Physics, 
School of Food and Nutritional Sciences, 
University of Shizuoka, 
Yada 52-1, Shizuoka 422-8526, Japan
 }

$\mbox{}^\ddag${\large
Department of Physics, Faculty of Education, Shizuoka University,
Shizuoka 422-8529, Japan
}
\end{center}

\vfill

{\large Abstract}\nl
A localized configuration is found
in the 5D bulk-boundary theory
on an $S_1/Z_2$ orbifold model of 
Mirabelli-Peskin.
A bulk scalar and the extra (fifth) component of
the bulk vector constitute the configuration. 
$\Ncal=1$ SUSY is preserved.
The effective potential of the SUSY theory is obtained
using the background field method. 
The vacuum is treated in a general way by allowing
its dependence on the extra coordinate.
Taking into account the {\it supersymmetric boundary condition}, 
the 1-loop full potential is obtained.
The scalar-loop contribution to the Casimir energy
is also obtained. Especially we find a {\it new} type which
depends on the brane
configuration parameters besides the $S_1$ periodicity
parameter.

\vspace{0.5cm}

PACS NO:
\ 11.10.Kk,
\ 11.27.+d,
\ 12.60.Jv,
\ 12.10.-g,
\ 11.25.Mj,
\ 04.50.+h 
\nl

\vspace{0.5cm}

{\bf 1}\ {\it Introduction}\q
Through the development of the 
recent several years, it looks that  
the higher-dimensional approach begins to obtain the citizenship 
as an important building tool in constructing a unified theory.
Among many ideas in this approach,  
the system of {\it bulk and boundary} theories becomes a fascinating model
of the unification. 
The boundary is regarded as
our 4D world. It is inspired by the M, string and D-brane theories\cite{HW96}.
One pioneering paper, giving a concrete field-theory realization, 
 is that by Mirabelli and Peskin\cite{MP97}. 
They consider 
5D supersymmetric Yang-Mills theory with a boundary matter.
The boundary couplings with the bulk world
are uniquely fixed by the SUSY requirement. 
They demonstrated some consistency of the bulk and boundary quantum effects
by calculating {\it self-energy} of the scalar matter field.
Here we examine the vacuum configuration and
the effective potential.

Contrary to the motivation of ref.\cite{MP97}, 
we do not seek the SUSY breaking mechanism, rather
we make use of the SUSY-invariance properties
in order to make the problem as simple as possible.
The SUSY symmetry is so restrictive that we only
need to calculate some small portion of
all possible diagrams.

In the calculation of the effective potential
of the 5D model, we recall that of the Kaluza-Klein model. 
The dynamics quantumly produces the effective
potential which describes the Casimir effect\cite{AC83,SI85}.
The situation, however, is 
different from the present case
in the following points:\ 
1)\ the 4D reduction mechanism;\ 
2)\ Z$_2$-symmetry;\ 
3)\ treatment of the vacuum with respect to
 the extra-coordinate dependence;\ 
4)\ supersymmetry;\ 
5)\ characteristic length scales.
We will compare the present result with the KK case.

{\bf 2}\ {\it Mirabelli-Peskin Model}\q
Let us consider the 5 dimensional flat space-time with the signature
(-1,1,1,1,1).
\footnote{
Notation is basically the same as ref.\cite{Hebec01}.
} 
The space of the fifth
component is taken to be ($S_1$), 
with the periodicity $2l$, and has the $Z_2$-orbifold condition.
\begin{eqnarray}
\xf\ra\xf+2l\ (\mbox{periodicity})\com\q
\xf\change -\xf\ (Z_2\mbox{-symmetry})\pr
\label{mp1b}
\end{eqnarray}
We take a 
5D bulk theory $\Lcal_{bulk}$ which is
coupled with a 4D matter theory $\Lcal_{bnd}$ on a "wall" at $\xf=0$
and with $\Lcal'_{bnd}$ on the other "wall" at $\xf=l$.
The boundary Lagragians are, in the bulk action,  described by
 the delta-functions along the extra axis $x^5$.
\begin{eqnarray}
S=\int d^5x\{\Lcal_{blk}+\del(x^5)\Lcal_{bnd}+\del(x^5-l){\Lcal'}_{bnd}
+\mbox{periodic part}\}
\pr\label{mp1}
\end{eqnarray}
We consider both bulk and boundary quantum effects.

The bulk dynamics is given by the 5D super YM theory
which is made of 
a vector field $A^M\ (M=0,1,2,3,5)$, 
a scalar field $\Phi$, 
a doublet of symplectic Majorana fields $\la^i\ (i=1,2)$, 
and a triplet of auxiliary scalar fields $X^a\ (a=1,2,3)$:
\begin{eqnarray}
\Lcal_{SYM}=-\half\tr {F_{MN}}^2-\tr (\na_M\Phi)^2
-i\tr(\labar_i\ga^M\na_M\la^i)
+\tr (X^a)^2+\tr (\labar_i[\Phi,\la^i])\com   
\label{mp2}
\end{eqnarray}
where all bulk fields are the {\it adjoint} representation
(its suffixes: $\al,\be,\cdots $)
of the gauge group $G$. 
The bulk Lagrangian $\Lcal_{SYM}$ 
is invariant under the 5D SUSY transformation.
This system has the symmetry of
8 real super charges.
As the 5D gauge-fixing term, we take the Feynman
gauge:
\begin{eqnarray}
\Lcal_{gauge}=-\tr (\pl_MA^M)^2=-\half (\pl_MA^M_{~\al})^2
\pr
\label{mp4b}
\end{eqnarray}
The corresponding ghost Lagrangian is given by
\begin{eqnarray}
\Lcal_{ghost}=-2\,\tr \pl_M\cbar\cdot \na^M(A)c
=-2\,\tr\pl_M\cbar\cdot (\pl^Mc+ig[A^M,c])
\com
\label{mp4c}
\end{eqnarray}
where $c$ and $\cbar$ are the complex ghost fields. 
We take the following bulk action.
\begin{eqnarray}
\Lcal_{blk}=\Lcal_{SYM}+\Lcal_{gauge}+\Lcal_{ghost}
\pr
\label{mp4d}
\end{eqnarray}

It is known that we can consistently project out $\Ncal=1$ SUSY
multiplet, which has 4 real super charges, 
by assigning $Z_2$-parity 
to all fields in accordance with the 5D SUSY. 
A consistent choice is given as:\  $P=+1$ for 
$A^m, \la_L, X^3$;  $P=-1$ for 
$A^5, \Phi, \la_R, X^1, X^2$ ($m=0,1,2,3$). 
Then ($A^m,\la_L,X^3-\na_5\Phi$) constitute
 an $\Ncal =1$ vector multiplet. 
Especially $\Dcal\equiv X^3-\na_5\Phi$ plays the role
of {\it D-field} on the wall. 
We introduce one 4 dim chiral multiplet ($\phi,\psi,F$) on the $\xf=0$ wall
and the other one ($\phi',\psi',F'$) on the $\xf=l$ wall: 
complex scalar fields $\phi,\phi'$, Weyl spinors $\psi,\psi'$, and
auxiliary fields of complex scalar $F,F'$. 
These are the simplest matter candidates and were taken
in the original theory\cite{MP97}. 
Using the $\Ncal=1$ SUSY property of the fields 
($A^m,\la_L,X^3-\na_5\Phi$),
we can find the following bulk-boundary coupling on the $\xf=0$ wall.
\begin{eqnarray}
\Lcal_{bnd}=-\na_m\phi^\dag \na^m\phi-\psi^\dag i\sibar^m \na_m\psi+F^\dag F
+\sqrt{2}ig(\psibar\labar_L\phi-\phi^\dag\la_L\psi)
+g\phi^\dag \Dcal\phi\com
\label{mp7}
\end{eqnarray}
where $    
\na_m\equiv \pl_m+igA_m,\ \Dcal =X^3-\na_5\Phi$.
We take the fundamental representation for $\phi,\phi^\dag$. 
The quadratic (kinetic) terms of the vector $A^m$, the gaugino spinor $\la_L$
and the 'auxiliary' field $\Dcal=X^3-\na_5\Phi$ are in the bulk world. 
In the same way we introduce the coupling between the matter fields
($\phi',\psi',F'$) on the $\xf=l$ wall and the bulk fields:\ 
$\Lcal'_{bnd}=(\phi\ra\phi', \psi\ra\psi', F\ra F' in \mbox{(\ref{mp7})})$. 
We note the interaction between the bulk fields and the boundary
ones is definitely fixed from SUSY.

{\bf 3}\ {\it SUSY Boundary Condition, Background Expansion and Generalized vacuum}\q
First we point out an important fact about
the SUSY effective potential. 
The 1-loop SUSY effective potential can be calculated
only by the scalar loop \footnote
{Non-scalar
{\it external} fields are always put zero from the definition
of the effective potential.
}
{\it up to the $F$- and $D$-independent terms}
in the off-shell treatment. 
If we trace the origin of this phenomenon, 
it is simply that
the auxiliary fields have the
{\it higher physical dimension} of $M^2$. They
cannot have the Yukawa coupling with fermions and vectors.
F and D-dependence in the SUSY effective
potential is very important to determine the vacuum behaviour. 
The above fact means that 
$dV^{eff}_{1-loop}/dD$ ( or $dV^{eff}_{1-loop}/dF$ )  
is definitely determined only by the scalar loop.
Miller\cite{Mil83PL,Mil83NP}, using the above fact, 
obtained  
F-tadpole or D-tadpole \cite{Wein73} 
(F and D-tadpole correspond to $dV^{eff}_{1-loop}/dF$
and $dV^{eff}_{1-loop}/dD$, respectively.) 
in general 4D SUSY theories. 
He noticed, if the theory preserve SUSY at the quantum level, 
the {\it $F$ and $D$-independent} parts in $V^{eff}_{1-loop}$ can be obtained, 
instead of calculating diagrams, 
by a {\it boundary condition} on the effective potential.
This is because, 
in the SUSY-preserving case, the effective potential
should satisfy:\ 
$V^{eff}(F=0,D=0)=0$ --{\it supersymmetric boundary condition}--.
He confirmed the correctness by comparing his results with
the results in the ordinary method. (See ref.\cite{AM98IJMPA}
for an application to unified models.)  
We follow Miller's idea.

Hence we may put, 
for the purpose of obtaining the 1-loop SUSY effective potential, 
the following conditions: 
\begin{eqnarray}
A^m=0\ (m=0,1,2,3)\com\q \la^i=\labar^i=0
\com\q \psi=0\com\q\psi'=0\com\q \la_L=0
\pr
\label{ep1}
\end{eqnarray}
Here the extra (fifth) component of the bulk vector $A^5$ 
does {\it not} taken to be zero because it is regarded
as a {\it 4D scalar} on the wall. 
The extra coordinate $\xf$ is regarded as {\it a parameter}.
Then $\Lcal_{blk}$    
reduces to
\begin{eqnarray}
\Lcal^{red}_{blk}[\Phi,X^3,A_5]
=\tr \left\{ -\pl_M\Phi\pl^M\Phi+X^3X^3-\pl_MA_5\pl^MA_5
+2g(\pl_5\Phi\times A_5)\Phi\right.\nn
\left. -g^2(A_5\times\Phi)(A_5\times\Phi)
-2\pl_M\cbar\cdot\pl^Mc-2ig\pl_5\cbar\cdot [A^5,c]\right\}
+\mbox{irrel. terms}
\com
\label{ep3}
\end{eqnarray}
where we have dropped terms of $2\tr X^1X^1=X^1_\al X^1_\al, 
2\tr X^2X^2=
X^2_\al X^2_\al$
as  'irrelevant terms' because they decouple from other fields. 
(Note $\tr (\pl_5\Phi\times A_5)\Phi
=(1/2)f_{\ab\ga}\pl_5\Phi_\al A_{5\be}\Phi_\ga$.) While
$\Lcal_{bnd}$, on the $\xf=0$ wall, reduces to
\begin{eqnarray}
\Lcal^{red}_{bnd}[\phi,\phi^\dag,X^3-\na_5\Phi]=-\pl_m\phi^\dag\pl^m\phi
+g(X^3_\al-\na_5\Phi_\al)\phi^\dag_\bep (T^\al)_{\bep\gap}\phi_\gap
+\mbox{irrl. terms} 
\com
\label{ep4}
\end{eqnarray}
where we have dropped $F^\dag F$-terms as the irrelevant terms.
$\alp, \bep$ are the suffixes of the fundamental representation.
In the same way, we obtain 
${\Lcal^{red}_{bnd}}'[\phi',{\phi'}^\dag,X^3-\na_5\Phi]$
on the $\xf=l$ wall.

Now we take the background-field method\cite{DeW67,tH73,IO82} 
to obtain the effective
potential. 
We expand all scalar fields ($\Phi, X^3, A_5; \phi, \phi'$), except ghosts, 
into the {\it quantum fields} (which are denoted again by the same symbols) 
and the {\it background fields}
($\vp, \chi^3, a_5; \eta, \eta'$).
\begin{eqnarray}
\Phi\ra\vp+\Phi\ ,\ 
X^3\ra \chi^3+X^3\ ,\ 
A_5\ra a_5+A_5\ ,\ \phi\ra\eta+\phi\ ,\ \phi'\ra\eta'+\phi'\ ,
\label{ep5}
\end{eqnarray}
We treat the ghosts $c$ and $\cbar$ as quantum fields. 

We state a new point in the present use of the background-field
method. 
Usually we take the following procedure in order 
to obtain the vacuum\cite{SI84}.

[{\it Ordinary} procedure of the vacuum search]\nl
1) First we obtain the effective potential
assuming the {\it scalar} property of the vacuum
(as described in (\ref{ep1}))
and the {\it constancy} of the scalar vacuum
expectation values. \nl
2) Then we take the minimum of the effective
potential.

In the present case, however, we have the {\it extra} coordinate $\xf$. 
We have "freedom" in the treatment of the vacuum expectation values
because $\xf$ is regarded as a simple {\it parameter}. 
We require that
{\it the background fields may be constant only in 4D world, not necessarily
in 5D world}. 
We may allow the background fields to depend on the extra coordinate $x^5$. 
This standpoint 
gives us an interesting possibility to the higher dimensional model and 
generalizes the vacuum of the system. 

When the background fields ($\vp, \chi^3, a_5; \eta, \eta'$) satisfy 
the {\it field equations} derived from (\ref{ep3}) and (\ref{ep4}),
we say they satisfy the {\it on-shell} condition. The equations are
, in the order of the variations $(\del\Phi_\al, \del A_{5\al}, \del\chi^3_\al, 
\del\phi^\dag_\alp, \del\phi'^\dag_\alp)$, 
respectively given as, 
\begin{eqnarray}
\pl_5Z_\al+g(Z\times a_5)_\al=0\com\q
{\pl_5}^2a_{5\al}-g(\vp\times Z)_\al=0\com\nn
\chi^3_\al+g\{ \del(\xf)\eta^\dag T^\al\eta
+\del(\xf-l)\eta'^\dag T^\al\eta'\}=0\com\nn
d_\be (T^\be\eta)_\alp=0\com\q
d_\be (T^\be\eta')_\alp=0\com\nn
\mbox{with the definition:\ }\qqq\qqq\qqq\qqq\nn
Z_\al\equiv -\pl_5\vp_\al+g(a_5\times\vp)_\al
-g\{ \del(\xf)\eta^\dag T^\al\eta+\del(\xf-l)\eta'^\dag T^\al\eta'\}\com\nn
d_\al\equiv \chi^3_\al-\na_5\vp_\al=
(\chi^3-\pl_5\vp+ga_5\times\vp)_\al
\com
\label{ep5b}
\end{eqnarray}
where we assume, based on the standpoint of the previous paragraph,
$\vp=\vp(\xf), \chi^3=\chi^3(\xf), a_5=a_5(\xf), \eta=\mbox{const},
\eta'=\mbox{const}$. 
The third equation guarantees $Z_\al=d_\al$. 
In the above derivation, we use the fact that 
total divergences, in the action, 
vanish from the {\it periodicity} condition. 
Because we seek the effective potential (an off-shell quantity), 
we generally do not need to assume the above on-shell condition. 
\footnote{
However the {\it minimum} of the effective potential should always
be consistent with the on-shell condition. 
The on-shell condition becomes important 
when we {\it restrict} the forms of the background fields.
(See later discussion.)
A new on-shell condition replace it. 
We should check that the new {\it minimum} 
is consistent with the new on-shell condition.}

The {\it quadratic} part w.r.t. the quantum fields 
($\Phi, X^3, A_5; \phi, \phi'$)
give us
the 1-loop quantum effect. This part 
is given as
\begin{eqnarray}
\Lcal^2_{blk}[\Phi,A_5,X^3]
=\tr\,\{ -\pl_M\Phi\pl^M\Phi
+ X^3 X^3 -\pl_M A_{5}\pl^M A_{5}\}\nn
+2g\,\tr\left[ (\pl_5\vp\times A_5)\Phi+(\pl_5\Phi\times a_5)\Phi
+(\pl_5\Phi\times A_5)\vp \right]\nn
-2g^2\tr\left[ (a_5\times\vp)(A_5\times \Phi)\right]
-g^2\tr (a_5\times\Phi+A_5\times\vp)^2\nn
-2\tr\,\{\pl_M\cbar\cdot\pl^Mc+ig\pl_5\cbar\cdot [a_5,c]\}\ ,\nn
\Lcal^2_{bnd}
=-\pl_m\phi^\dag\pl^m\phi
+g d_\al\phi^\dag T^\al\phi-ig^2[A_5,\Phi]_\al\eta^\dag T^\al\eta
\nn
+g(X^3_\al-\pl_5\Phi_\al-ig[a_5,\Phi]_\al-ig[A_5,\vp]_\al)
(\eta^\dag T^\al\phi+\phi^\dag T^\al\eta)\ ,\nn
{\Lcal^2_{bnd}}'=\{ \phi\ra \phi',\ \eta\ra\eta'\ \mbox{in }\Lcal^2_{bnd}\}
\com
\label{ep6}
\end{eqnarray}
where $d_\al= \chi^3_\al-\na_5\vp_\al$ 
 is the background (4 dimensional) D-field and 
$\phi^\dag T^\ga\phi\equiv \phi^\dag_\alp(T^\ga)_{\alp\bep}\phi_\bep$.
Now we can integrate out the auxiliary field $X^3_\al$ in
$\Lcal^2_{blk}+\del(x^5)\Lcal^2_{bnd}+\del(x^5-l){\Lcal^2_{bnd}}'$.  We obtain
the final "1-loop Lagrangian", necessary for the present purpose, as
\begin{eqnarray}
S^{(2)}[\Phi,A_5;\phi]=\int d^5X \left[
\Lcal_{blk}^2|_{X^3=0}-\del(x^5)\pl_m\phi^\dag \pl^m\phi\right.\nn
\left. +\del(x^5)\{
gd_\al (\phi^\dag T^\al\phi)-g\pl_5\Phi_\al 
(\eta^\dag T^\al\phi+\phi^\dag T^\al\eta)
-\frac{g^2}{2}\del(0)(\eta^\dag T^\al\phi+\phi^\dag T^\al\eta)^2
            \}                \right]
\com\label{ep9}
\end{eqnarray}
where $\del(x^5-l)$ part is dropped because we need not to consider
the quantum propagation in the $\xf=l$ brane.
\footnote{
The effect of the $\xf=l$ brane is in non-trivial background solutions
(vacuum configurations) derived by (\ref{ep5b}). It quantumly appears
in the effective potential as the present quantum effect. See the following
description.
}


{\bf 4}\ {\it Mass-Matrix and the Localized Background Configuration}\q
We are now ready for the full ( with respect to the coupling order) 
calculation of the 1-loop 
(we call this "1-loop full") 
effective potential. 
The "1-loop action" can be expressed as
\begin{eqnarray}
S^{(2)}=S^{ghost}+S^{free}+\int d^5X\nn
\times\half \left(\begin{array}{cccc}
\phi^\dag_{\al'} & \phi_{\al'} & \Phi_\al & A_{5\al} 
              \end{array}\right)
       {\left(\begin{array}{cc}
\left(\begin{array}{cc}
M_{\phi^\dag\phi} & M_{\phi^\dag\phi^\dag}  \\
M_{\phi\phi} & M_{\phi\phi^\dag}  \\
\end{array}
\right)_{\alp\bep}         &
\left(\begin{array}{cc}
M_{\phi^\dag\Phi} & 0 \\
 M_{\phi\Phi} & 0 
\end{array}
\right)_{\alp\be}          \\
\left(\begin{array}{cc}
M_{\Phi\phi} & M_{\Phi\phi^\dag} \\
0 & 0  
\end{array}
\right)_{\al\bep}         &
\left(\begin{array}{cc}
M_{\Phi\Phi}  & M_{\Phi A_5} \\
M_{A_5\Phi} & M_{A_5 A_5}
\end{array}
\right)_{\al\be}
             \end{array}\right)}
%
\left(\begin{array}{c}
\phi_{\be'} \\ \phi^\dag_{\be'} \\ \Phi_\be \\ A_{5\be}
              \end{array}\right)  ,\nn
S^{ghost}=-\int d^5X\left[
\pl_M\cbar_\al\cdot\pl^Mc_\al
+igf_{\ab\ga}\pl_5\cbar_\al\cdot a_{5\be}c_\ga
\right]\q\nn
S^{free}=\intdX
\left[ \tr\,\{ -\pl_M\Phi\pl^M\Phi
-\pl_M A_{5}\pl^M A_{5}\}-\del(\xf)\pl_m\phi^\dag\pl^m\phi\right]
,\label{det1}
\end{eqnarray}
where $S^{ghost}$ is decoupled from others, and the components $M'$s
are read from (\ref{ep9}). 

Now we restrict the form of the background fields 
in the present 5D approach. The relevant scalars
are $a_5$ and $\vp$ in the bulk. 
We should
take into account the {\it $x^5$-dependence}
and the Z$_2$-property of the background fields.

(i) Brane-anti-brane solution\nl
We take the following forms of $a_5(\xf)$ and $\vp(\xf)$,
which describe the localized (around $x^5=0$) configurations and
a natural generalization of the ordinary treatment stated before.
\begin{eqnarray}
a_{5\ga}(\xf)=\abar_\ga\,\ep (\xf)\com\q
\vp_\ga(\xf)=\vpbar_\ga\ep (\xf)\nn
\ep(\xf)=\left\{
\begin{array}{cc}
+1 & \mbox{for }2nl<\xf<(1+2n)l \\
0  & \mbox{for }\xf=nl \\
-1 & \mbox{for }(2n-1)l<\xf<2nl \end{array}
\right.\q n\in {\bf Z}\pr
\label{det6}
\end{eqnarray}
where $\ep(\xf)$ is the {\it periodic sign function}
with the periodicity $2l$.
\footnote{
We define the values at $\xf=nl$ to be $0$ in (\ref{det6})
in order to make the function $\ep(\xf)$ piece-wise continuous
and also to make it Fourier expandable.
}
 $\abar_\ga$ and $\vpbar_\ga$ are 
some positive constants. See Fig.1.
\begin{figure}
\centerline{ \psfig{figure=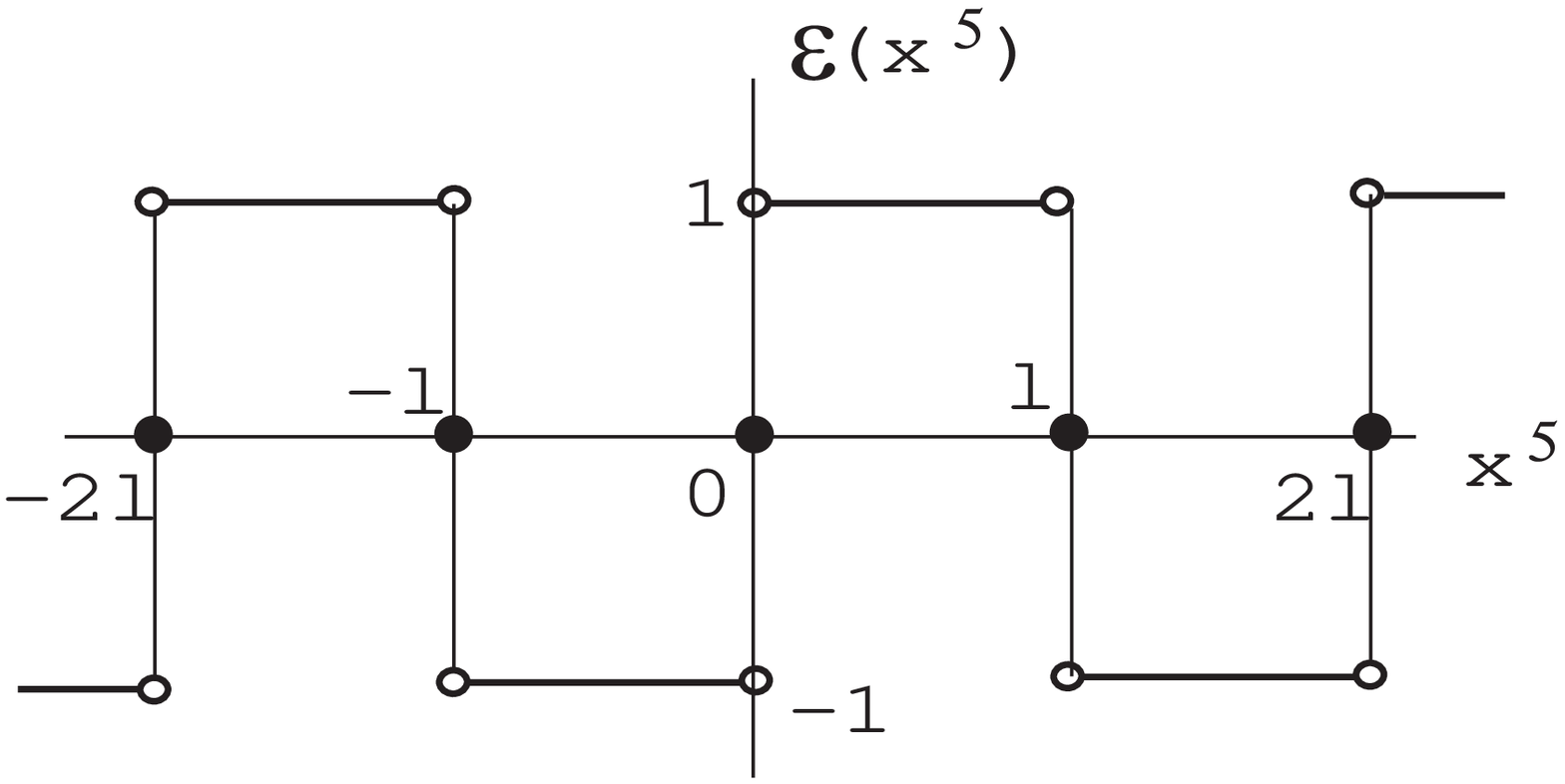,height=3cm,angle=0}}
\caption{ 
The graph of the periodic sign function $\ep(\xf)$, (\ref{det6}).
Background fields $a_5$ and $\vp$ behave as
$a_{5\ga}(\xf)=\abar_\ga\,\ep (\xf)\com\q
\vp_\ga(\xf)=\vpbar_\ga\ep (\xf)$.
}
\label{fig:Sign}
\end{figure}
It is the {\it thin-wall limit} of a (periodic) kink solution
and shows the {\it localization} of the fields. 

The background fields, (\ref{det6}),
satisfy the required boundary condition. 
We show they also {\it satisfy the on-shell condition} (\ref{ep5b}) 
for an appropriate
choice of $\abar, \vpbar, \eta, \eta'$ and $\chi^3$.
The assumed background forms are summarized as 
\begin{eqnarray}
\vp_\al(\xf)=\vpbar_\al\ep(\xf)\com\q a_{5\al}(\xf)=\abar_\al\ep(\xf)\com\nn
\eta_\alp=\mbox{const}\com\q \eta'_\alp=\mbox{const}\com\q
d_\al=\chi^3_\al-\na_5\vp_\al=\mbox{const}
\com
\label{C1}
\end{eqnarray}
where "const"'s mean some constants which generally may be different.
\footnote{
Although $\Dcal_\al$ is made of the bulk fields, it behaves as a boundary
field (D-field of $\Ncal=1$ SUSY multiplet), hence we consider
the case that its background value $d_\al$ is independent of $\xf$. 
} 
We note the relation
\begin{eqnarray}
\pl_5\vp_\ga=2\vpbar_\ga\{\del(\xf)-\del(\xf-l)\}
\com
\label{det15}
\end{eqnarray}
where $\del(\xf)$ is the {\it periodic} delta function with
the periodicity $2l$. 
The above equation expresses the {\it localization} of the bulk scalar
at $\xf=0$ and $\xf=l$. It is considered to be the field theoretical
version of the brane-anti-brane configuration. 
See Fig.2. 
\begin{figure}
\centerline{ \psfig{figure=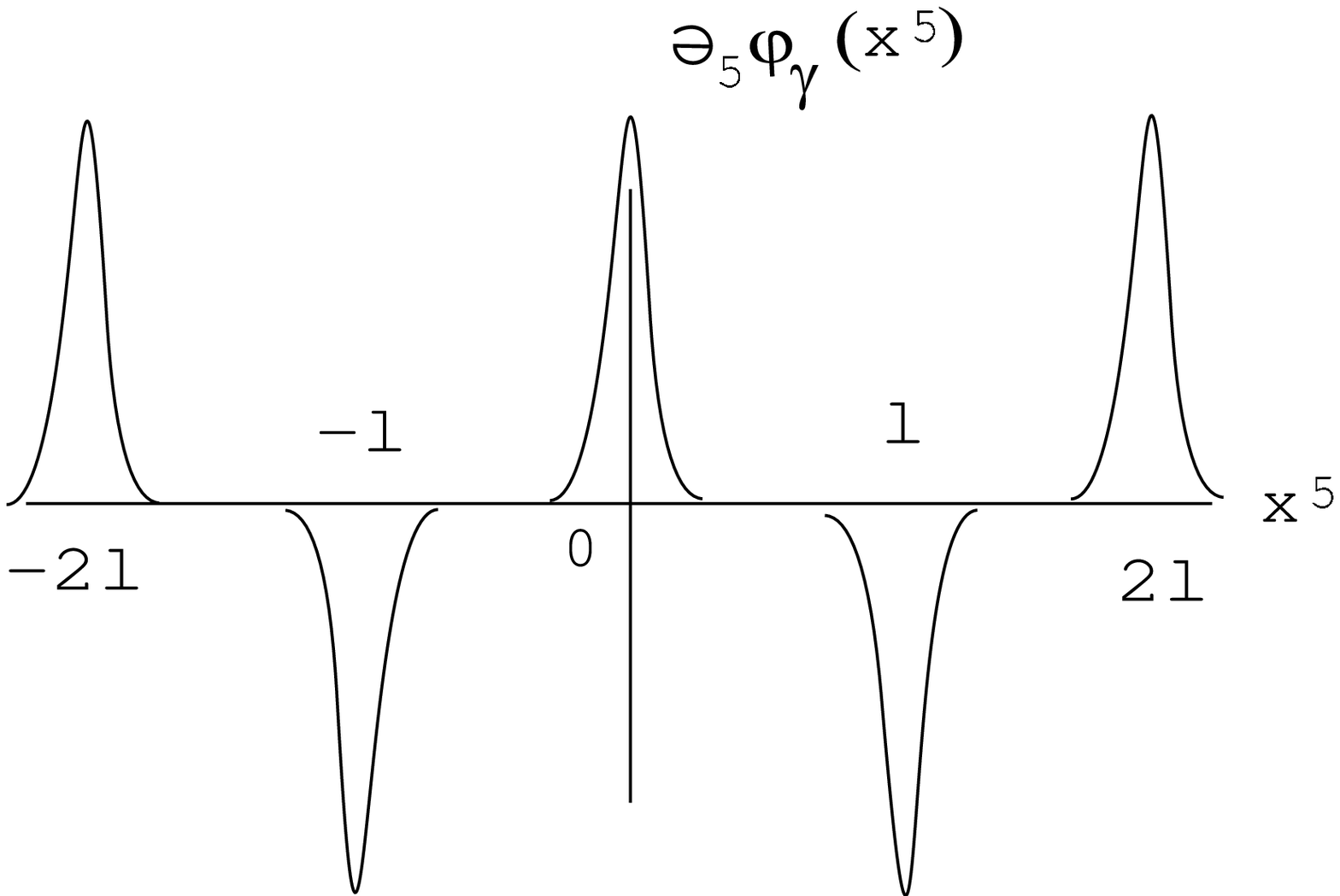,height=3cm,angle=0}}
\caption{ Behaviour of 
 $\pl_5\vp_\ga(x^5)$.
}
\label{fig:DelFunc}
\end{figure}
Using this relation, 
the first two equations of (\ref{ep5b}) are replaced by
\begin{eqnarray}
\pl_5\{ -\vpbar_\al\pl_5\ep+g\ep^2(\abar\times\vpbar)_\al
-g(\del(\xf)\eta^\dag T^\al\eta+\del(\xf-l)\eta'^\dag T^\al\eta')\}\nn
+g\ep \{(-\vpbar\pl_5\ep+g\ep^2(\abar\times\vpbar)
-g(\del(\xf)\eta^\dag T\eta+\del(\xf-l)\eta'^\dag T\eta'))
\times\abar\}_\al=0\ ,\nn
\abar_\al{\pl_5}^2\ep-g\ep\{\vpbar\times 
(-\vpbar\pl_5\ep+g\ep^2(\abar\times\vpbar)
-g(\del(\xf)\eta^\dag T\eta+\del(\xf-l)\eta'^\dag T\eta'))\}_\al=0 
\ .
\label{C2}
\end{eqnarray}
We note here the following things.
\begin{enumerate}
\item
When $\abar_\al\propto\vpbar_\al$, the following relations hold:\ 
$(\abar\times\vpbar)_\al=f_{\ab\ga}\abar_\be\vpbar_\ga=0$.
\item
We may use the equation:\ 
$\pl_5(\del(\xf)-\del(\xf-l))\times\mbox{const}=0$, 
in the field equation on condition that
the arbitrary variation $\del A^5_\al(\xf)$,
which is used to derive the second equation of (\ref{ep5b}), satisfies
the relation: $\pl_5(\del A^5_\al)|_{\xf=0}=\pl_5\del(\del A^5_\al)|_{\xf=l}$.
\footnote{ 
See the next footnote.
}
\item
$\ep(\xf)^2=1$, 
$\ep(\xf)^3=\ep(\xf)$, 
$\pl_5(\ep(\xf))=2(\del(\xf)-\del(\xf-l))$,
$\half\pl_5\{\ep(\xf)^2\}=(\del(\xf)-\del(\xf-l))\ep(\xf)=0$
.
\end{enumerate}
Then we can conclude that (\ref{C1}) is a {\it solution of the field
equation (\ref{ep5b})} for the following choice.
\begin{eqnarray}
\frac{1}{c} \abar_\al=\vpbar_\al
=-\frac{g}{2}\eta^\dag T^\al\eta=\frac{g}{2}\eta'^\dag T^\al\eta'\com\q
\chi^3_\al=-g(\del(\xf)-\del(\xf-l))\eta^\dag T^\al\eta
\com
\label{C3}
\end{eqnarray}
where $c$ is a free parameter. 
\footnote{
A special choice, $c=0$, is given by :\ 
$
\abar_\al= 0,\ 
\vpbar_\al
=-\frac{g}{2}\eta^\dag T^\al\eta
=\frac{g}{2}\eta'^\dag T^\al\eta'\ ,
\chi^3_\al=-g(\del(\xf)-\del(\xf-l))\eta^\dag T^\al\eta
$.
This solution does not require the item 2 below eq.(\ref{C2}).
}
In this choice $d_\al=0$ is concluded. 
Hence the final two equations of (\ref{ep5b}) are satisfied. 
We can regard these as the new on-shell condition due to the
restriction of the background fields (\ref{det6}). 
The present vacuum (minimum point of the effective potential)
should be consistent with (\ref{C3}). 

(ii) Sawtooth-wave solution

We consider another solution. 
\begin{eqnarray}
a_{5\ga}(\xf)=\abar_\ga\times [\xf]_p\com\q
\vp_\ga(\xf)=\vpbar_\ga \times [\xf]_p\com  \nn
 \mbox{[$\xf$]$_p$} =
\left\{
\begin{array}{cc}
\xf & -l<\xf <l \\
0   &   \xf=l \\
\mbox{periodic} & \mbox{other regions} 
\end{array}
\right.\com
\label{noko1}
\end{eqnarray}
where $[\xf]_p$ is the {\it sawtooth-wave} (periodic linear function)
with the periodicity $2l$. $\abar_\ga$ and $\vpbar_\ga$ are 
some positive constants. See Fig.3.
\begin{figure}
\centerline{ \psfig{figure=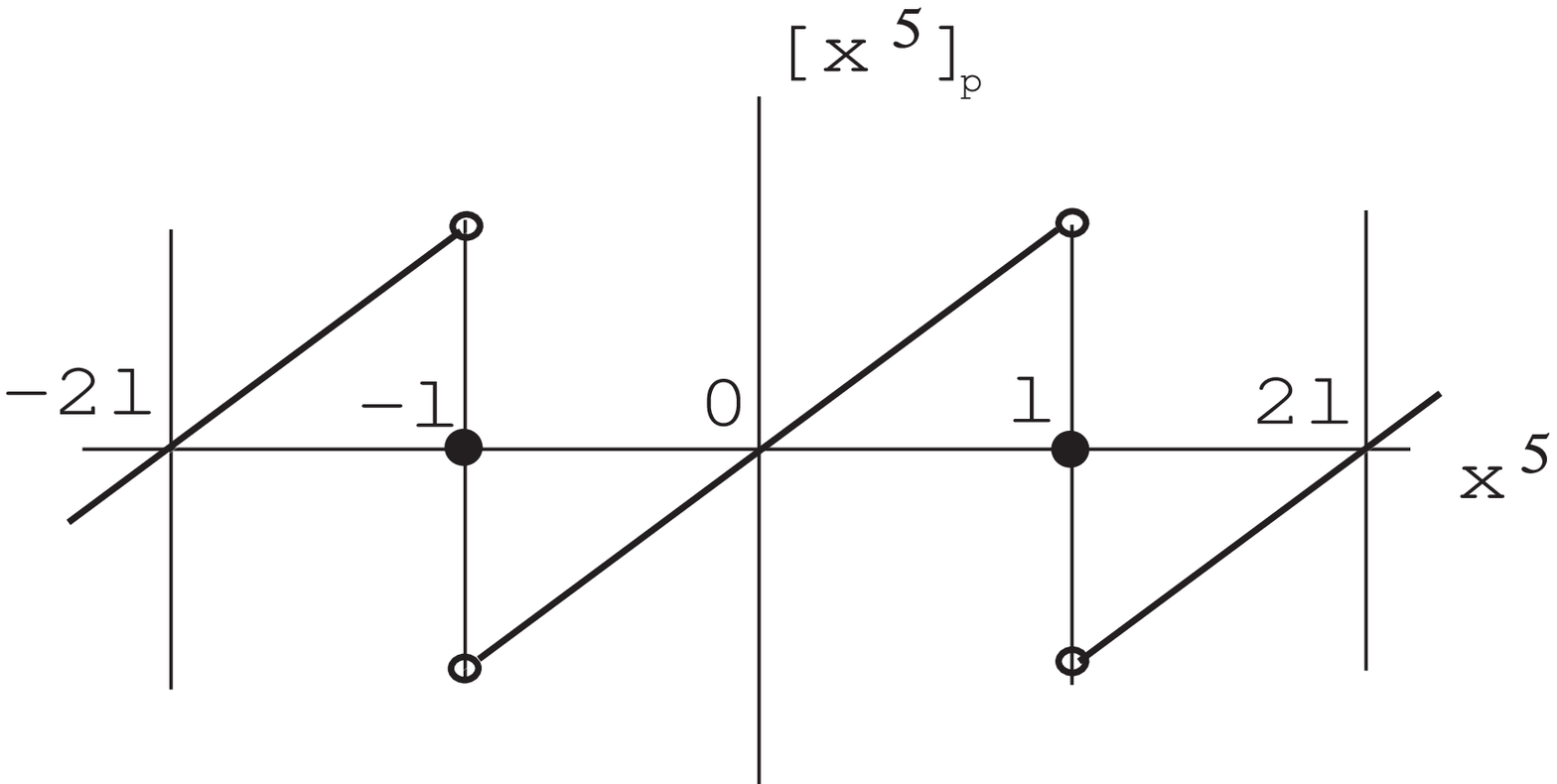,height=3cm,angle=0}}
\caption{ 
The graph of the sawtooth wave $[\xf]_p$, (\ref{noko1}).
Background fields $a_5$ and $\vp$ behave as
$a_{5\ga}(\xf)=\abar_\ga\times [\xf]_p\com\q
\vp_\ga(\xf)=\vpbar_\ga\times [\xf]_p$.
}
\label{fig:SawTooth}
\end{figure}
Using (\ref{noko1}), with the following relations in $-l<\xf\leq l$:\ 
$\pl_5\vp_\al=\vpbar_\al-2l\vpbar_\al\del(\xf-l), 
{\pl_5}^2\vp_\al=-2l\vpbar_\al\del'(\xf-l); 
\pl_5a_{5\ga}=\abar_\ga-2l\abar_\ga\del(\xf-l), 
{\pl_5}^2a_{5\ga}=-2l\abar_\ga\del'(\xf-l)$, 
we can find a solution in the following way.
First we consider, as in the previous solution, the case that the two scalars
$\abar_\al$ and $\vpbar_\al$ are "parallel" in the
isospace:\  $\abar_\al=\mbox{const}\times\vpbar_\al$. 
Then the key quantity $Z_\al$ can be written as
\begin{eqnarray}
Z_\al=d_\al=-\vpbar_\al \{1-2l\del(\xf-l)\}
-g\{ \del(\xf)\eta^\dag T^\al\eta+\del(\xf-l)\eta'^\dag T^\al\eta' \}
\pr
\label{noko1b}
\end{eqnarray}
Now we require that $d_\al$ should be independent of the extra
axis $\xf$. Then we obtain 
\begin{eqnarray}
\eta_\alp=\eta^\dag_\alp=0\com\q
\vpbar_\al=\frac{g}{2l}\eta'^\dag T^\al\eta'=\frac{1}{c}\abar_\al\com\q
Z_\al=d_\al=-\vpbar_\al
\com
\label{noko1c}
\end{eqnarray}
where $c$ is a free parameter. 
The first equation of (\ref{ep5b}) is satisfied. 
The second equation requires:\ 
$\abar_\al{\pl_5}^2[\xf]_p=-2l\abar_\al\pl_5(\del(\xf-l))=0$. 
It means the variation $\del A_{5\al}$, which is used
to derive the second equation, should satisfy the {\it Neumann
boundary} condition:\ 
\begin{eqnarray}
\frac{\pl}{\pl\xf}\del A_{5\al}|_{\xf=l}=0
\pr
\label{noko2}
\end{eqnarray}
(For a special case $c=0$ ($a_{5\al}=0$), the above
condition is not necessary.) 
The third equation gives
$\chi^3_\al=-g\del(\xf-l)\eta'^\dag T^\al\eta'$. 
The fourth equation of (\ref{ep5b}) is satisfied. 
The fifth equation gives 
the condition on 
the values of $\eta'_\alp$:\ 
\begin{eqnarray}
d_\be (T^\be\eta')_\alp=-\frac{g}{2l}
(\eta'^\dag T^\be\eta')(T^\be\eta')_\alp=0
\pr
\label{noko3}
\end{eqnarray}
All on-shell conditions are satisfied by the above choice.
Especially, $d_\al=-\vpbar_\al=-\frac{g}{2l}\eta'^\dag T^\al\eta'$. 
From the form of $\pl_5\vp_\al=\vpbar_\al-2l\vpbar_\al\del(\xf-l)$ 
(see Fig.4), 
these backgrounds are considered to describe 
the mixture of a non-localized and a localized (at one end)
configurations. The form of the sawtooth-wave solution (Fig.3)
is reminiscent of the AdS$_5$ solution of the dilaton
in the Randall-Sundrum model although the latter one is Z$_2$ even
whereas the present one is Z$_2$ odd.

\begin{figure}
\centerline{ \psfig{figure=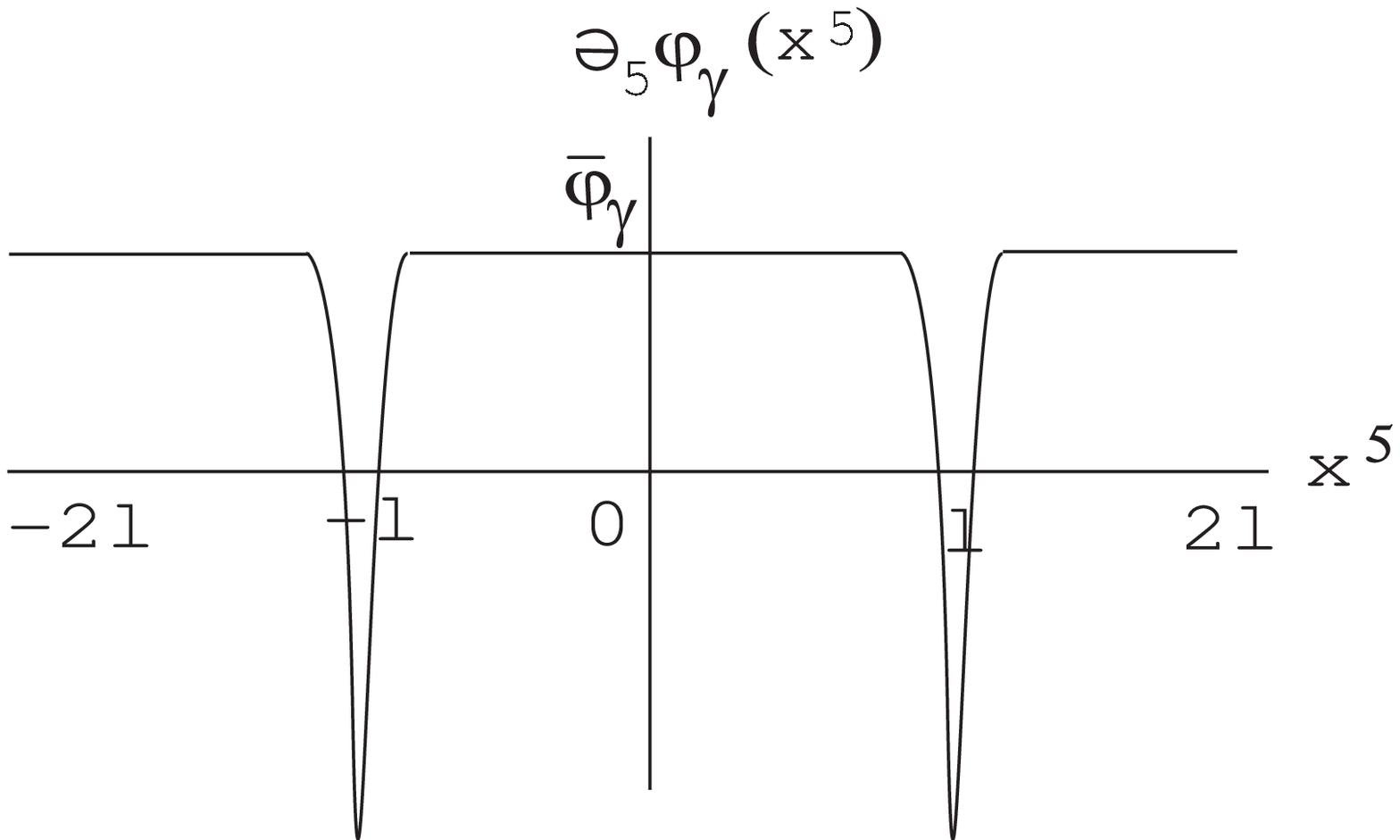,height=3cm,angle=0}}
\caption{ Behaviour of 
 $\pl_5\vp_\ga(x^5)$ for the sawtooth wave solution (\ref{noko1}).
}
\label{fig:Const}
\end{figure}

Taking the localized solution (i), 
we evaluate $S^{(2)}$, (\ref{det1}), furthermore. 
\footnote{
The solution (ii) will be treated in a forthcoming paper.
}
 From the periodicity ($\xf\ra\xf+2l$) and the Z$_2$ 
property, the bulk quantum fields $\Phi(X), A_5(X)$ and $c(X)$
can be KK-expanded as
\begin{eqnarray}
\Phi(x,\xf)=
\frac{1}{\sql}\sum_{n=1}^{\infty}\Phi_n(x)\sin(\npl\xf)\com\q
A_5(x,\xf)=
\frac{1}{\sql}\sum_{n=1}^{\infty}A_n(x)\sin(\npl\xf)\com\nn
c(x,\xf)=
\frac{1}{2\sql}\left\{ c_0(x)+2\sum_{n=1}^{\infty}c_n(x)\cos(\npl\xf)
               \right\}\pr
\label{det3}
\end{eqnarray}
(The $Z_2$-parity of the ghost field is even because it should be
the same as that of 
the gauge parameter $\La$\ : $\del A^M=\pl^M\La-ig[A^M,\La]$. )
Now we use 
the Fourier expansion of the periodic sign function,
\begin{eqnarray}
\ep(x)=\frac{4}{\pi}\sum_{n=0}^\infty\frac{1}{2n+1}
\sin\{\frac{(2n+1)\pi}{l}x\}
\com
\label{det9}
\end{eqnarray}
and the relation: 
\begin{eqnarray}
\int_{-l}^{l}d\xf\ep(\xf)\cos(\mpl\xf)\sin(\npl\xf)
=-\frac{2l}{\pi}Q_{mn}\com\nn
Q_{mn}=\left\{
\begin{array}{cc}
\frac{1}{m-n}\q & m-n=\mbox{odd} \\
0    &  m-n=\mbox{even}
\end{array}
\right.
\pr
\label{det10}
\end{eqnarray}
Noting the above equations and (\ref{det15}), 
we can express $S^{(2)}$ in terms of the 4D integral as follows.
\begin{eqnarray}
S^{(2)}=S^{ghost}+\int d^4x\times\nn
\half \left(\begin{array}{cccc}
\phi^\dag_{\al'} & \phi_{\al'} & \Phi_{m\al} & A_{m\al} 
              \end{array}\right)
       {\left(\begin{array}{cc}
\left(\begin{array}{cc}
\Mcal_{\phi^\dag\phi} & \Mcal_{\phi^\dag\phi^\dag}  \\
\Mcal_{\phi\phi} & \Mcal_{\phi\phi^\dag}  \\
\end{array}
\right)_{\alp\,\bep}         &
\left(\begin{array}{cc}
\Mcal_{\phi^\dag\Phi} & 0 \\
\Mcal_{\phi\Phi} & 0 
\end{array}
\right)_{\alp \, n\be}          \\
\left(\begin{array}{cc}
\Mcal_{\Phi\phi} & \Mcal_{\Phi\phi^\dag} \\
0 & 0  
\end{array}
\right)_{m\al \, \bep}         &
\left(\begin{array}{cc}
\Mcal_{\Phi\Phi}  & \Mcal_{\Phi A} \\
\Mcal_{A\Phi} & \Mcal_{A A}
\end{array}
\right)_{m\al \, n\be}
             \end{array}\right)}
%
\left(\begin{array}{c}
\phi_{\be'} \\ \phi^\dag_{\be'} \\ \Phi_{n\be} \\ A_{n\be}
              \end{array}\right)  ,
\label{det19}
\end{eqnarray}
where the integer suffixes $m$ and $n$ runs from 1 to $\infty$, and 
each component is described as
\begin{eqnarray}
\Mcal_{\phi^\dag_\alp\phi_\bep}=
\pl^2\del_{\alp\bep}+
gd_\ga (T^\ga)_{\alp\bep}-g^2\del(0)(T^\ga\eta)_\alp
(\eta^\dag T^\ga)_\bep\com\nn
\Mcal_{\phi^\dag_\alp\phi^\dag_\bep}=
-g^2\del(0)(T^\ga\eta)_\alp (T^\ga\eta)_\bep\com\q
\Mcal_{\phi_\alp\phi_\bep}= 
-g^2\del(0)(\eta^\dag T^\ga)_\alp (\eta^\dag T^\ga)_\bep\com\nn
\Mcal_{\phi_\alp\phi^\dag_\bep}= 
\pl^2\del_{\alp\bep}+
gd_\ga (T^\ga)_{\bep\alp}-g^2\del(0)(\eta^\dag T^\ga)_\alp (T^\ga\eta)_\bep
\com\nn
\Mcal_{\phi^\dag_\alp\Phi_{n\be}}=
-\frac{g}{\sql}(T^\be\eta)_\alp\npl=\Mcal_{\Phi_{n\be}\phi^\dag_\alp}\com\q
\Mcal_{\phi_\alp\Phi_{n\be}}=
-\frac{g}{\sql}(\eta^\dag T^\be)_\alp\npl=\Mcal_{\Phi_{n\be}\phi_\alp}\com\nn
\Mcal_{\Phi_{m\al}\Phi_{n\be}}=
-\{-\pl^2+(\npl)^2\}\del_{mn}\del_\ab
-g^2f_{\al\del\tau}f_{\be\ga\tau}\abar_\del\abar_\ga\del_{mn}
+\frac{4g}{l}f_{\ab\ga}\abar_\ga mQ_{mn}\com\nn
\Mcal_{\Phi_{m\al} A_{n\be}}=
g^2f_{\ab\tau}f_{\ga\del\tau}\abar_\ga\vpbar_\del\del_{mn}
-g^2f_{\ga\al\tau}f_{\be\del\tau}\abar_\ga\vpbar_\del\del_{mn}
-\frac{2g}{l}f_{\ab\ga}\vpbar_\ga mQ_{mn}
=\Mcal_{A_{n\be}\Phi_{m\al}} ,\nn
\Mcal_{A_{m\al} A_{n\be}}=
-\{-\pl^2+(\npl)^2\}\del_{mn}\del_\ab
-g^2 f_{\al\ga\tau}f_{\be\del\tau}\vpbar_\ga\vpbar_\del\del_{mn}\com
\label{det20}
\end{eqnarray}
where the kinetic (free) part is also included 
($\pl^2\equiv \pl_m\pl^m$) 
in the ``Mass'' matrix and the repeated indices imply the Einstein's 
summation convention. 
$S^{ghost}$ is decoupled and is given by
\begin{eqnarray}
S^{ghost}=\int d^4x\left\{
\half\pl_m\cbar_{0\al}\pl^mc_{0\al}
+\sum_{k=1}^{\infty}\left(\pl_m\cbar_{k\al}\pl^mc_{k\al}
-(\frac{k\pi}{l})^2\cbar_{k\al}c_{k\al}\right)
\right.\nn
+
\left.
\sum_{n=1}^\infty\sum_{k=1}^\infty\cbar_{n\al}(x)
[-\frac{2ig}{l}f_{\al\ga\be}\abar_\ga nQ_{nk}]c_{k\be}(x)
\right\}
\pr
\label{det21}
\end{eqnarray}
This contribution is treated independently from others.

{\bf 5}\ {\it Effective Potential of Bulk-Boundary System}\q
The effective potential
is obtained from the eigen values of the mass-matrix obtained
in (\ref{det19}),(\ref{det20}) and (\ref{det21}). We examine
the behaviour for two typical cases.

(A)$\eta=0,\eta^\dag=0$(Bulk-Boundary decoupled case)\nl
We look at the potential from the vanishing scalar-matter point.
In this case the singular terms, 
$\del(0)$-terms, disappear and the matrix $\Mcal$ decouples to the boundary part 
($\phi,\phi^\dag$) 
and the bulk part ($\Phi, A$). The former part gives the following eigen values.
\begin{eqnarray}
\la_\pm=-k^2\pm\frac{g}{2}\sqrt{d^2}\com\q
d^2\equiv {d_1}^2+{d_2}^2+{d_3}^2
\com\q k^2=k_mk^m\com
\label{B5}
\end{eqnarray}
where we take G=SU(2) and the doublet representation for the boundary matter
fields.  $k^m$ is the 4D momentum. 
This gives, taking the supersymmetric boundary condition,
the following potential before the renormalization:
\begin{eqnarray}
V^{eff}_{1-loop}
=\int\frac{d^4k}{(2\pi)^4}
\ln \{1-\frac{g^2}{4}\frac{d^2}{(k^2)^2}\}
=-\frac{g^2}{4}\int\frac{d^4k}{(2\pi)^4}\frac{d^2}{(k^2)^2}
+O(g^4)
\com
\label{B7}
\end{eqnarray}
The last perturbative (w.r.t. $g$ ) form 
is logarithmically divergent. It can be checked by the perturbative
 calculation. 
It is renormalized
by the {\it bulk} wave function of $X^3$ and $\Phi$.  Here the 4D world's 
connection to the Bulk world appears. The quantum fluctuation
within the boundary influence the bulk world through the renormalization.
The form of (\ref{B7}) is similar to the 4D super QED\cite{Mil83NP}. 
We see the present model produces a desired effective potential on the brane.

The bulk part of $\Mcal$ and the ghost part do {\it not} depend on the field $d$.
They and their eigenvalues depend only on 
the brane parameters, $\abar$ and $\vpbar$, and the size of the extra space, $l$.
In the SUSY boundary condition, their contribution to the vacuum energy
is zero. The scalar loop contribution is expected to be cancelled by 
the quantum effect of the non-scalar fields.
Let us, however, examine the scalar-loop contribution to the Casimir energy (potential). General case is technically difficult. We consider the
{\it large circle limit}:\ 
$\gh^2\equiv \frac{g^2}{l}=\mbox{fixed}\ll 1\ ,\ 
\ahat=\sqrt{l}\abar=\mbox{fixed}\ ,\ 
\vphat=\sqrt{l}\vpbar=\mbox{fixed}\ ,\ 
l\ra\infty $.
This is the situation where the circle is large
compared with the inverse of the domain wall height.
($\ahat$ and $\vphat$ have the dimension of $M$. )
We notice, in this limit, $Q_{mn}$-terms disappear. In the
"propagator" terms of the bulk quantum fields, 
KK-mass terms $m^2\pi^2/l^2$
disappear. All KK-modes equally contribute to the vaccum energy.
The eigen values of the bulk part of $\Mcal$ can be
easily obtained. 
In particular, for the special case $\ahat=0$, the nontrivial
factor is only $k^2+\gh^2\vphat^2$. Hence each KK-mode
equally contributes to the vacuum energy as
\begin{eqnarray}
V^{eff}_{1KK-mode}
\propto\int\frac{d^4k}{(2\pi)^4}
\ln \{1+\gh^2\frac{\vphat^2}{k^2}\}
\com
\label{B7.16}
\end{eqnarray}
This quantity is quadratically divergent. 
After an appropriate normalization,
the final form should become, based on the dimensional analysis, 
the following one.
\begin{eqnarray}
V^{eff}_{1-loop}
=\gh^2(c_1\frac{\vphat^2}{l^2}+c_2\frac{\ahat^2}{l^2}
+c_3\frac{\ahat\cdot\vphat}{l^2})+O(\gh^4)
\com
\label{B7.17}
\end{eqnarray}
where $c_1, c_2$ and $c_3$ are some finite constants which are calculable
after we know the bulk quantum dynamics sufficiently.
This is a {\it new} type Casimir energy.
This is the reason why we have examined the scalar-loop contribution. 
Comparing the ordinary one (\ref{B10b}) explained soon,
it is new in the following points:\ 
1) it depends on the brane parameters $\vphat$ and $\ahat$
besides the extra-space size $l$;\ 
2) it depends on the gauge coupling $\gh$;\ 
3) it is proportional to $1/l^2$.

We expect the above result of Casimir energy are cancelled 
by the spinor and vector-loop contribution in the present SUSY theory.
The unstable Casimir potential do {\it not} appear in SUSY theory.

(B)$\abar=0,\vpbar=0$\nl
In this case, $Q_{mn}$-terms disappear and we do have
no localized (brane) configuration. 
The bulk background configuration is trivial:\ $a_5(\xf)=0,\ \vp(\xf)=0$. 
5D bulk quantum fields
fluctuate 
with the periodic boundary condition in the extra space. 
This is similar to the 5D Kaluza-Klein case mentioned
in the introduction. 
The eigen values for the bulk part,
$c(X), \cbar(X), \Phi(X)$ and $A_5(X)$ are commonly given by, 
\begin{eqnarray}
\la_n=-k^2-(\npl)^2\com\q n=1,2,3,\cdots
\com
\label{B10}
\end{eqnarray}
The eigen values are basically the same as the KK case \cite{AC83}.
They depend {\it only} on the radius (or the periodicity) parameter $l$.
It gives the scalar-loop contribution to the Casimir potential. From the
dimensional analysis, after the renormalization, it has the
following form.
\begin{eqnarray}
\frac{1}{l}V^{eff}_{scalar}= \frac{\mbox{const}}{l^5}
\pr
\label{B10b}
\end{eqnarray}
We expect again this contribution is cancelled 
by the spinor and vector fields. 

The eigenvalues for the boundary part is obtained as 
a complicated expression involving the following terms:
\begin{eqnarray}
S\equiv\eta^\dag\eta\ ,\ d^2=d_\al d_\al\ ,\ 
d\cdot V\equiv d_\al \, \eta^\dag T^\al\eta\ ,\ 
V^2\equiv (\eta^\dag T^\al\eta)^2 
\pr
\label{B20a}
\end{eqnarray}
We have the full expression in the computer file.
In the manipulation of eigen-values search
(determinant calculation), 
we face
the following combination of terms.
\begin{eqnarray}
\del(0)+\frac{1}{l}\sum_{m=1}^{\infty}
\frac{(\pi m/l)^2}{-\la-k^2-(\pi m/l)^2}
\pr
\label{B19}
\end{eqnarray}
The first term comes from the {\it singular} terms
in $\Mcal$, the second from the KK-mode sum.
Using the relation$\sum_{m\in\bfZ}1=2l\del(0)$, 
the above sum leads to a {\it regular} quantity.
\begin{eqnarray}
{\del(0)}|_{sm}=
\frac{1}{2l}\sum_{m\in\bfZ}
\frac{\la+k^2}{\la+k^2+(\pi m/l)^2}
=\left\{
\begin{array}{cc}
\half\sqrt{\la+k^2}\coth\{l\sqrt{\la+k^2}\} & \la>-k^2 \\
\half\sqrt{-\la-k^2}\cot\{l\sqrt{-\la-k^2}\} & -k^2>\la 
\end{array}
\right.
\pr
\label{B20}
\end{eqnarray}
We have confirmed this "smoothing" phenomenon
occurs at the 1-loop full level.

For some interesting cases, we present the explicit forms
of the eigenvalues.\nl
(i) $\eta=\eta^\dag=0\ (d\cdot V=0, V^2=0, S=0)$\nl
This is a special case of (A), the decoupled case.
\begin{eqnarray}
\la_1=\la_2=\la_+\com\q\la_3=\la_4=\la_-\com\q
\la_\pm=-k^2\pm \frac{g}{2}\sqrt{d^2}
\pr
\label{B22}
\end{eqnarray}
It is consistent with Case (A).

(ii) $d\cdot V\neq 0$, others=$0$ ($S=0, d^2=0, V^2=0$)\nl
Interesting eigenvalues come from the solutions of the
following equation.
\begin{eqnarray}
\q (\la+k^2)^2-\frac{g^3}{2}d\cdot V\frac{\sqrt{\la+k^2}}{2}
\coth l\sqrt{\la+k^2}=0
\pr
\label{B23}
\end{eqnarray}
To confirm the correctness, we look at the perturbative
aspect of this 1-loop full result. 
First expanding the above expression by $1/k^2$
(propagator expansion), and then taking the terms
up to the 1st order w.r.t. $g^2/l$, we obtain
\begin{eqnarray}
(\la+k^2)^2-\frac{g^3}{4}d\cdot V\sqrt{k^2}
\coth l\sqrt{k^2}=0
\pr
\label{B25}
\end{eqnarray}
Two eigenvalues $\la_1,\la_2$ satisfy
\begin{eqnarray}
\la_1\la_2=
(k^2)^2\left( 1-\frac{g^3}{4}d\cdot V
\frac{\sqrt{k^2}\coth l\sqrt{k^2}}{(k^2)^2}\right)
\pr
\label{B26}
\end{eqnarray}
This result is consistent with the perturbative result
(the vertex correction on the boundary)
 up to the order of $g^3$. 
The full-order eigenvalues, the solutions of (\ref{B23}),
correspond to the 1-loop {\it full} effective potential.
 
{\bf 6}\ {\it Conclusion}\q
We have analyzed the effective potential of 
the Mirabelli-Peskin model. 
The explicit forms are obtained for some cases.
An interesting
localized configuration (solution) is found 
in the bulk scalar
and the extra-component of the bulk vector
when we solve the field equation (on-shell condition).
The vacuum is generalized in connection with
the treatment of the extra axis. 
We treat $\xf$ as a parameter which is 
independent of the 4D world. 
The important role of the D-field,
$\Dcal_\al=X^3_\al-\na_5(A)\Phi_\al$,
in the 4D world is confirmed. 
In this SUSY invariant theory, the Casimir
force vanishes. Its scalar-loop contribution
is obtained from the explicit matrix
elements depending on the boundary parameters
$\abar$, $\vpbar$ and $l$. 
Besides the ordinary type, 
we find a {\it new} type form
of the Casimir energy which is characteristic for
the brane model. When SUSY is broken in some mechanism,
the new type potential could become an important
distinguished quantity of the bulk-boundary system
from the ordinary KK system.

We hope the present result improves 
the understanding of the quantum dynamics of 
the bulk-boundary system.

\begin{flushleft}
{\bf Acknowledgment}
\end{flushleft}
The authors thank N. Sakai for valuable comments 
when this work, still at the primitive stage, 
was presented at the Chubu Summer School 2002 (Tsumagoi, Gunma, Japan,
2002.8.30-9.2). A part of
this work was done when 
one of the author (S.I.) stayed at
 DAMTP(Univ. of Cambridge,2002.11.22-2003.2.10). He thanks 
G.W. Gibbons and G. Silva for comments and discussions. 
The hospitality there is acknowledged. 
He also thanks the governor of the Shizuoka prefecture for
the financial support.

\vs 1


\end{document}